\documentclass[journal]{IEEEtran}

\ifCLASSINFOpdf

\else

\fi

\usepackage[utf8]{inputenc}
\usepackage{amssymb}
\usepackage{stfloats}
\usepackage{placeins}
\usepackage{physics}
\usepackage{overpic}
\usepackage{caption}
\usepackage{amsmath}
\usepackage{comment}
\usepackage{mathtools}
\usepackage{booktabs}
\usepackage{cite}
\usepackage{subfig}
\usepackage{lipsum}
\usepackage{graphicx}
\usepackage{textcomp,gensymb}
\usepackage{glossaries}
\usepackage{multicol}
\usepackage{algorithm}
\usepackage[noend]{algpseudocode}
\usepackage{tikz}
\usepackage[english]{babel}
\usepackage{amsthm}
\theoremstyle{plain}
\usepackage{dirtytalk}

\graphicspath{ {./images/} }

\usepackage{amsmath}
\DeclareMathOperator*{\argmax}{arg\,max}

\usepackage{mathtools}

\usepackage{tabularx}
\newcolumntype{L}[1]{>{\raggedright\arraybackslash}p{#1}}
\newcolumntype{C}[1]{>{\centering\arraybackslash}p{#1}}
\newcolumntype{R}[1]{>{\raggedleft\arraybackslash}p{#1}}
\usepackage{multirow}
\usepackage{mathtools, cuted}

\addto\captionsenglish{}
\newcommand{\AM}[1]{\textcolor{black}{#1}}
\newacronym{fec}{FEC}{forward error correction}
\newacronym{3gpp}{3GPP}{3rd Generation Partnership Project}
\newacronym{6g}{6G}{sixth-generation}    
\newacronym{iot}{IoT}{Internet of Things}
\newacronym{ntn}{NTN}{Non-Terrestrial Network}
\newacronym{leo}{LEO}{Low Earth Orbit}
\newacronym{geo}{GEO}{Geosynchronous Earth Orbit}
\newacronym{isl}{ISL}{Inter-Satellite Link}
\newacronym{gsl}{GSL}{Ground-to-Satellite Link}
\newacronym{qos}{QoS}{Quality of Service}
\newacronym{ofdma}{OFDMA}{Orthogonal Frequency-Division Multiple Access}
\newacronym{b5g}{B5G}{beyond 5G}
\newacronym{mimo}{MIMO}{Multiple-Input Multiple-Output}
\newacronym{embb}{eMBB}{Enhanced Mobile Broadband}
\newacronym{urllc}{URLLC}{ultra-reliable and low-latency communications}
\newacronym{mmtc}{mMTC}{massive machine-type communications}
\newacronym{ue}{UE}{User Equipment}
\newacronym{nr}{NR}{New Radio}
\newacronym{gap}{GAP}{Generalized Assignment Problem}
\newacronym{mgap}{MGAP}{Multi-Level Generalized Assignment Problem}
\newacronym{csi}{CSI}{channel state information}    
\newacronym{ran}{RAN}{radio access network}
\newacronym{5g}{5G}{the 5th generation of mobile networks}
\newacronym{uav}{UAV}{unmanned aerial vehicle}
\newacronym{ap}{AP}{access point}
\newacronym{sic}{SIC}{successive interference cancellation}
\newacronym{mmimo}{mMIMO}{massive \gls{mimo}}
\newacronym{snr}{SNR}{signal-to-noise ratio}
\newacronym{sinr}{SINR}{signal-to-interference plus noise ratio}
\newacronym{fifo}{FIFO}{first-in first-out}
\newacronym{mab}{MAB}{multi-armed bandit}
\newacronym{rl}{RL}{reinforcement learning}
\newacronym{noma}{NOMA}{non-orthogonal multiple access}
\newacronym{rv}{RV}{random variable}
\newacronym{drl}{DRL}{deep \gls{rl}}
\newacronym{irsa}{IRSA}{irregular repetition slotted ALOHA}
\newacronym{oma}{OMA}{orthogonal multiple access}
\newacronym{fdma}{FDMA}{frequency division multiple access}
\newacronym{tdma}{TDMA}{time division multiple access}
\newacronym{mdp}{MDP}{Markov decision process}
\newacronym{bs}{BS}{base station}
\newacronym{rs}{RS}{rate-splitting}
\newacronym{rsma}{RSMA}{rate-splitting multiple access}
\newacronym{dtmc}{DTMC}{discrete-time Markov chain}
\newacronym{cscg}{CSCG}{circularly symmetric complex Gaussian}
\newacronym{ack}{ACK}{acknowledgement}
\newacronym{nack}{NACK}{negative ACK}
\newacronym{aoi}{AoI}{age-of-information}
\newacronym{tare}{TRE}{time-averaged reconstruction error}
\newacronym{tacae}{TCAE}{time-averaged cost of actuation error}
\newacronym{udc}{UC}{update-delivery cost}
\newacronym{sc}{SC}{superposition coding}
\newacronym{dof}{DoF}{degrees-of-freedom}
\newacronym{siso}{SISO}{single-input single-output}
\newacronym{crlb}{CRLB}{Cram\'{e}r-Rao lower bound}
\newacronym{isa}{ISA}{isolated spectral allocation}
\newacronym{reir}{REIR}{radar estimation information rate}
\newacronym{dir}{DIR}{data information rate}
\newacronym{isac}{ISAC}{integrated sensing and communication}
\newacronym{v2x}{V$2$X}{vehicle-to-everything}
\newacronym{awgn}{AWGN}{additive white Guassian noise}
\newacronym{pdf}{PDF}{probability density function}
\newacronym{fim}{FIM}{Fisher information matrix}

\begin{document}
\title{Coexistence of Radar and Communication with Rate-Splitting Wireless Access}

\author{\IEEEauthorblockN{Anup~Mishra, \IEEEmembership{Member, IEEE}, Israel Leyva-Mayorga, \IEEEmembership{Member, IEEE} and Petar~Popovski, \IEEEmembership{Fellow, IEEE}\vspace{-0.8cm}}

\thanks{The authors Anup Mishra, Israel Leyva-Mayorga and Petar Popovski are with the Department of Electronic Systems, Aalborg University, Aalborg 9220, Denmark (e-mail:anmi@es.aau.dk; ilm@es.aau.dk; petarp@es.aau.dk).}}

\maketitle
\begin{abstract}
Future wireless networks are envisioned to facilitate the seamless coexistence of communication and sensing functionalities, thereby enabling the much-touted \gls{isac} paradigm. A key challenge in \gls{isac} is managing inter-functionality interference while maintaining a balanced performance trade-off. In this work, we propose a \gls{rs}-inspired approach to address this challenge in an uplink \gls{isac} scenario, where a \gls{bs} serves an uplink communication user while detecting a radar target. We derive inner bounds on ergodic data information rate for communication user and the ergodic radar estimation information rate for sensing target. A closed-form solution is also derived for the optimal power split in \gls{rs} that maximizes the communication user’s performance. Compared to \gls{oma}- and \gls{noma}-inspired approaches, the proposed approach achieves a more favorable sensing-communication trade-off {by virtue of the decoding order flexibility introduced through splitting the communication message}. Notably, this is the first work to employ a \gls{rs}-inspired strategy as a general framework for non-orthogonal coexistence of sensing and communication, extending its applicability beyond traditional digital-only settings.
\end{abstract}
\glsresetall
 \vspace{-0.1cm}
\glsresetall
 \vspace{-0.4cm}
\section{Introduction}\label{Intro}
\IEEEPARstart{F}{uture} wireless networks such as \gls{b5g} and \gls{6g} systems are gearing up to embrace sensing functionality\cite{Masouros@ISaC,Yuanwei@NOMA_ISaC}. With growing similarities in radio resources, hardware platforms, and signal processing techniques between communication and sensing, their integration has attracted significant attention from academia and industry alike\cite{Masouros@ISaC,Yuanwei@Uplink_ISaC}. This \gls{isac} paradigm promises to enable myriads of use-cases, including smart cities, remote sensing, \gls{iot}, and \gls{v2x} connectivity, among others\cite{Yuanwei@NOMA_ISaC,Yuanwei@Uplink_ISaC}. \AM{A central challenge in \gls{isac}, however, lies in effectively managing both inter-user interference within the communication functionality and inter-functionality interference between communication and sensing, while achieving a favorable trade-off between the two\cite{Masouros@ISaC, Yuanwei@NOMA_ISaC}. In the existing literature, the former is addressed using \textit{multiple access-assisted schemes}, which leverage classical multiple access strategies such as \gls{oma} and \gls{noma} to suppress inter-user interference within the communication domain\cite{Yuanwei@NOMA_ISaC}. In multi-user \gls{isac} scenarios, such suppression implicitly improves joint performance by enhancing the communication functionality \cite{NOMA_Assisted@Downlink,Yuanwei@NOMA_ISaC}. On the other hand, \textit{multiple access-inspired schemes} explicitly target {inter-functionality interference}, i.e., interference between the communication and sensing functionalities. These schemes operate either via spectral isolation or through spectral sharing with \gls{sic}, and are commonly referred to as \gls{oma}-inspired and \gls{noma}-inspired approaches, respectively\cite{Bliss@InnerBound,Yuanwei@Uplink_ISaC,NOMA_Inspired@Downlink}}. Notably, while multiple access-assisted schemes have been thoroughly investigated in the \gls{isac} literature, research into multiple access-inspired schemes are in their early stages\cite{Yuanwei@Uplink_ISaC,Yuanwei@NOMA_ISaC}. Bridging this gap is crucial to addressing interference challenges, enhancing resource efficiency, and enabling robust \gls{isac} capabilities in future wireless networks\cite{Masouros@ISaC,Mishra@tutorialRSMA}.
\begin{figure}
    \centering
    \includegraphics[width=0.75\linewidth]{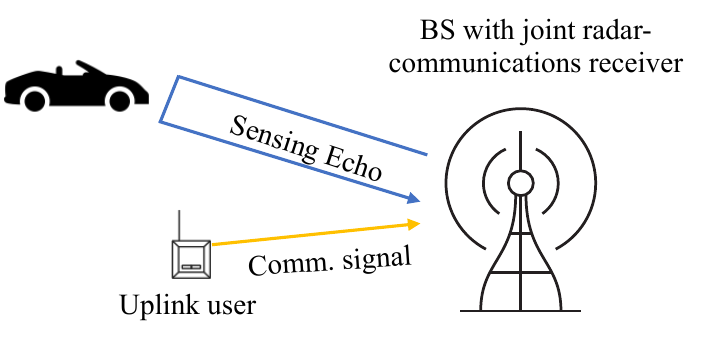}
    \caption{A basic setup with a \gls{bs} serving a communication user in the uplink and simultaneously sensing a radar target.}
    \label{fig:Schematic}\vspace{-0.7cm}
\end{figure}
\par {{Existing multiple access-inspired schemes inherit certain limitations from their communication counterparts\cite{Yuanwei@NOMA_ISaC}}. Specifically, the \gls{oma}-inspired approach may suffer from low resource efficiency, while the \gls{noma}-inspired approach may lead to one functionality being interference-limited by the other due to its fixed decoding order} \cite{Yuanwei@NOMA_ISaC}. To overcome these limitations, the \gls{rs} approach offers a promising alternative\cite{Mishra@tutorialRSMA}. Within the communication domain, \gls{rs} has gained significant traction as a robust interference management technique\cite{RSintro16bruno,mishra2021ratesplitting,Mishra@tutorialRSMA}. In \gls{rs}, a user's message is split into multiple parts, transmitted using \gls{sc} at the transmitter and decoded via \gls{sic} at the receiver\cite{RSintro16bruno}. Building on \gls{rs}, the \gls{rsma} scheme caters to multi-user scenarios by optimizing message splitting and power allocation \cite{Mishra@tutorialRSMA}. This enables RSMA to manage inter-user interference in the communication domain effectively by partially decoding interference and partially treating it as noise, thereby outperforming conventional multiple access schemes\cite{Mishra@tutorialRSMA}. \AM{Notably, RSMA-assisted designs have been employed in \gls{isac} to achieve improved performance trade-offs over \gls{oma} and \gls{noma}-assisted designs\cite{Longfei@isac,RSMA_Assisted@Downlink}. However, to the best of our knowledge, an \gls{rs}-inspired approach specifically targeting inter-functionality interference in \gls{isac} has not yet been investigated in the existing literature.} 
\par Motivated by the above, in this work, we investigate the performance bound of \gls{rs}-inspired approach in a joint sensing-communication system comprising an active, mono-static, pulsed radar and a communication user. Interestingly, just as \gls{rsma} generalizes \gls{noma} for communication, the \gls{rs}-inspired approach here generalizes the \gls{noma}-inspired approach. To this end, we consider a \gls{bs} with a joint radar-communication receiver serving a communication user in the uplink and simultaneously sensing a radar target; see Fig.~\ref{fig:Schematic}. This joint receiver is capable of simultaneously estimating radar target parameters from the radar echoes and decoding the received communication signals\cite{Bliss@InnerBound,Yuanwei@NOMA_ISaC}. We also assume that the radar system operates without any constraints on the maximum unambiguous range. On the other hand, the communication user is employing \gls{rs} at its side to transmit information. Building on this, we derive the sensing-communication co-existence performance bounds for \gls{rs}. For the radar target, the performance bound is measured in terms of ergodic \gls{reir}\cite{Bliss@InnerBound,Yuanwei@Uplink_ISaC}, whereas for the communication user it is ergodic \gls{dir}\cite{Bliss@InnerBound,Yuanwei@NOMA_ISaC}. Moreover, we utilize the derived bounds to obtain the optimal power split for \gls{rs} that maximizes the \gls{dir} of the communication user. {We demonstrate that \gls{rs}-inspired approach achieves a superior performance trade-off compared to conventional \gls{oma} and \gls{noma} inspired approaches\cite{Yuanwei@NOMA_ISaC}. To our best knowledge, this is the first \gls{rs}-inspired work for \gls{isac} that takes forward the concept of \gls{rs} beyond digital signals only}, and puts it as a general method for including non-orthogonal access for sensing signals, thereby providing a systematic and parametrized approach to effectuate non-orthogonal sensing and communication waveforms.
\vspace{-0.1cm}
\section{System Model}\label{Sysmod}
To begin with, we outline the key assumptions of this work, which align with those in \cite{Bliss@InnerBound}, a study that examines the performance bounds of conventional multiple access- inspired approaches. While the analysis in this paper focuses on range estimation, primarily to enable performance comparison with prior work, the proposed approach can be easily extended to estimate other parameters \cite{Bliss@InnerBound,Yuanwei@NOMA_ISaC}. However, the work done in this paper can be extended to other estimation parameters as well\cite{Bliss@InnerBound}. Subsequently, the assumptions are:
\begin{enumerate}
    \item Based on prior observations (since the target is being tracked), the \gls{bs} is able to accurately estimate the target cross-section and predict the range up to some error, which has a Gaussian distribution\cite{Yuanwei@Uplink_ISaC,Bliss@InnerBound}.
    \item Only the time portion where the radar return and communication signals overlap is considered for analysis\cite{Bliss@InnerBound}.
\end{enumerate}
 Next, we delineate the signal model for the radar return of the target, and the communication user.
 \vspace{-0.3cm}
\subsection{Signal model}
We consider a system operating in the complex-baseband. The \gls{bs} transmits a radar signal $r\left(t\right)$ with power $P_{r}$ and unit variance\cite{Bliss@InnerBound}. \AM{We assume ideal self-interference suppression, following~\cite{Bliss@InnerBound,Yuanwei@NOMA_ISaC,Yuanwei@Uplink_ISaC}.  As a first approximation, non-ideal interference suppression increases the noise floor; however, a detailed modeling of the impact of residual self-interference is out of the scope for this letter.} Moreover, the \gls{bs} sets the configuration parameters for the  communication user e.g., pulse shapes, power and rate split for \gls{rs}, etc. For the radar signal, we denote the complex combined antenna gain, radar cross-section, \AM{propagation loss} by $a_{r}$, and time-delay by $\tau_{r}$.  
\par Next, since the communication user employs \gls{rs} at the transmitter, its message is split into two parts \cite{Mishra@tutorialRSMA}. \AM{The two parts will allow for flexible interference management between the communication and sensing functionalities by strategically ordering the communication streams relative to the sensing signal.} Subsequently, the two parts are independently encoded into streams of unit variance, $s_{c,1}\left(t\right)$ and $s_{c,2}\left(t\right)$. These streams are allocated powers $P_{c,1}$ and $P_{c,2}$ respectively, such that $P_{c,1}+P_{c,2}\leq P_{c}$, \AM{where $P_{c}$ represents the} total uplink transmit power at the user. While we consider one communication user here, the work done in this paper can be extended to multiple users case. Subsequently, the transmit signal of the communication user is expressed as\cite{Mishra@tutorialRSMA,RSintro16bruno}
\begin{equation}\label{eq:comm_tx}
    x\left(t\right)=\sqrt{P_{c,1}}\,{s}_{c,1}\left(t\right) + \sqrt{P_{c,2}}\,{s}_{c,2}\left(t\right).
\end{equation}
The complex combined antenna gain and propagation loss for  $x\left(t\right)$ is denoted by $b_{c}$. \AM{The propagation losses $a_r$ and $b_c$ account for both large-scale fading and small-scale fading effects, and are assumed to be known at the \gls{bs}\cite{Bliss@InnerBound,Yuanwei@Uplink_ISaC}.} Thereafter, at the \gls{bs}, the overall signal received by the joint radar-communication receiver is passed through a brick-wall filter matched to the bandwidth $B$ of the system\cite{Bliss@InnerBound}. \AM{This joint radar-communications complex-baseband received signal, denoted by $y\left(t\right)$, is given by\cite{Bliss@InnerBound}}
\begin{equation}\label{eq:joint_rx_signal}
    y\left(t\right)=b_{c}{x}\left(t\right) + \sqrt{P_{r}}\,a_{r}{r}\left(t-\tau_{r}\right)+n\left(t\right),
\end{equation}
where ${n}\left(t\right)$ is the \gls{awgn} with variance $\sigma_{n}^{2}=\kappa_{B}T_{\textrm{temp}}B$. Here, $\kappa_{B}$ and $T_{\textrm{temp}}$ denote the Boltzman constant and effective temperature, respectively. At the \gls{bs}, the joint radar-communication receiver first decodes one stream of the communication user, then the radar return signal, and finally the other communication stream.  \AM{Without loss of generality, we assume the decoding order to be ${s}_{c,1}\left(t\right)\rightarrow {r}\left(t-\tau_{r}\right) \rightarrow {s}_{c,2}\left(t\right)$.}
\begin{figure}[!b]
    \centering \vspace{-0.6cm}
    \includegraphics[width=\linewidth]{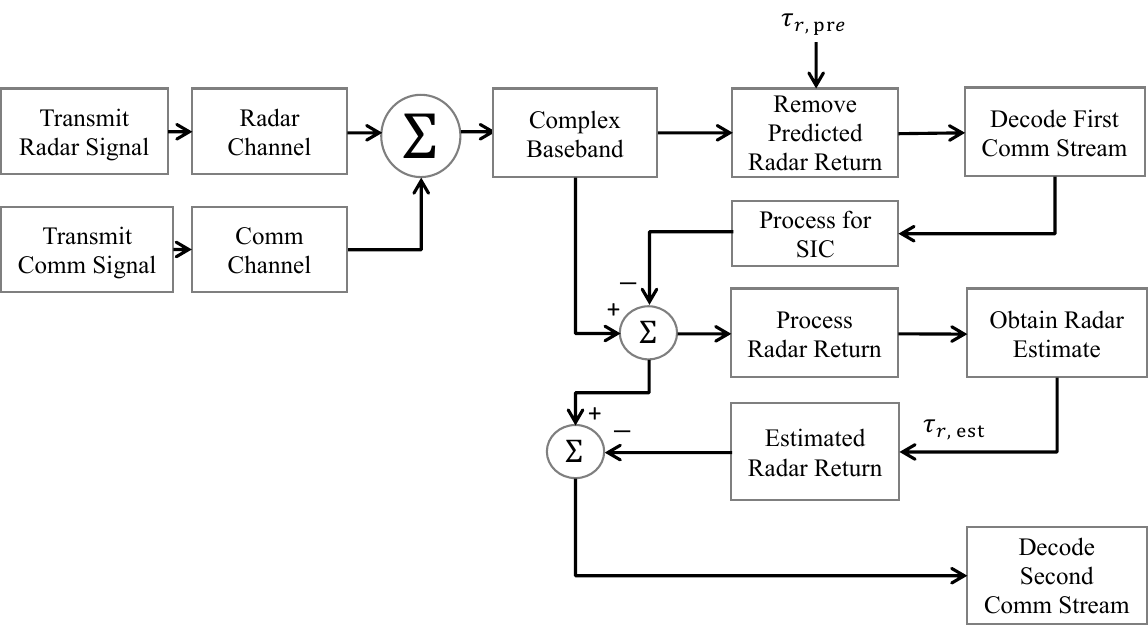}
    \caption{Joint radar-communications system block diagram with \gls{rs} at the communication user and \gls{sic} at the \gls{bs}.}
\label{fig:RS_SIC_Framework}
\end{figure}
\AM{The processing of the received baseband signal requires: 1) decoding the first communication stream with the \emph{predicted} radar return time-delay; 2) estimating the radar time-delay; and 3) decoding the second communication stream with the \emph{estimated} radar return. Such a decoding order is distinct from {multiple access-assisted} approaches, which aim to mitigate interference among communication users, typically treating the sensing signal as noise \cite{Yuanwei@NOMA_ISaC,Yuanwei@Uplink_ISaC}. In contrast, the proposed approach explicitly targets {inter-functionality interference}, necessitating a fundamental reinterpretation of both the decoding and estimation processes. This shift in objective introduces non-trivial analytical challenges, including the design of optimal power allocation, selection of decoding order, and derivation of the \gls{crlb} for radar estimation. Finally, the proposed approach also {generalizes the \gls{noma}-inspired approach}, obtained by disabling one of the communication streams.} Fig.~\ref{fig:RS_SIC_Framework} illustrates the framework for the \gls{rs}-inspired approach.
 \vspace{-0.25cm}
\subsection{First Communication Stream With Predicted Radar Return Suppressed}
As noted in Section~\ref{Sysmod}, the radar target is tracked, and its predicted delay $\tau_{r,\textrm{pre}}$ is known, with Gaussian fluctuation $n_{\tau_{r},\textrm{proc}}$ due to process noise~\cite{Bliss@InnerBound,Yuanwei@NOMA_ISaC}. Thus, $\tau_r = \tau_{r,\textrm{pre}} + n_{\tau_{r},\textrm{proc}}$, with variance $\sigma_{\tau_{r},\textrm{proc}}^2 = \mathbb{E}{|n_{\tau_{r},\textrm{proc}}|^2}$. This prediction is used to subtract the radar echo from $y(t)$ when decoding the first communication stream~\cite{Bliss@InnerBound,Yuanwei@Uplink_ISaC}. Subsequently, we get
\begin{equation}
\begin{split}
    \Tilde{y}\left(t\right)=&\,  b_{c}\sqrt{P_{c,1}}\,{s}_{c,1}\left(t\right) + b_{c}\sqrt{P_{c,2}}\,{s}_{c,2}\left(t\right)+\\
    &\sqrt{P_{r}}\,a_{r}\left[{r}\left(t-\tau_{r}\right)-{r}\left(t-\tau_{r,\textrm{pre}}\right)\right]+n\left(t\right).  
\end{split}
\end{equation}
\AM{The interference plus noise power from the first communications stream's point of view is given by\cite{Bliss@InnerBound,Yuanwei@Uplink_ISaC}}
\begin{equation}
    \sigma_{\textrm{int}+\textrm{n},1}^{2}= \abs{b_{c}}^{2}P_{c,2} + P_{r}\abs{a_{r} }^{2}\gamma^{2}B^{2}\sigma_{\tau_{r},\textrm{proc}}^{2} +\sigma_{n}^{2},
\end{equation}
\AM{where $\gamma^{2}=\left(2\,\pi\right)^{2}/12$ for a flat spectral wave\cite{Bliss@InnerBound}. Subsequently, the \gls{bs} decodes $s_{c,1}(t)$ with interference from ${r}\left(t-\tau_{r}\right)$ and stream $s_{c,2}(t)$. Thereafter, using \gls{sic}, the \gls{bs} removes $s_{c,1}(t)$ to estimate $\tau_{r}$, and decode stream $s_{c,2}(t)$.}
\vspace{-0.5cm}
\subsection{Radar Return Signal With Second Communication Stream As Interference}
In this subsection, we derive the \AM{\gls{crlb}} for the time-delay estimation of the radar signal. To this end, we consider a simple matched filter or correlation receiver for time-delay estimation\cite{Bliss@InnerBound}. \AM{Assuming perfect \gls{sic} of the first communication stream $s_{c,1}$; as considered in \cite{Bliss@InnerBound,Yuanwei@Uplink_ISaC} for deriving the bounds for \gls{noma}-inspired approach}; the received signal at the time-delay estimator is expressed as 
\begin{equation}\label{eq:Radar_Sig}
    z\left(t\right)= a_{r}\sqrt{P_{r}}\,{r}\left(t-\tau_{r}\right) + b_{c}\sqrt{P_{c,2}}\,{s}_{c,2}\left(t\right) + n\left(t\right).
\end{equation}
\AM{Note that the radar receiver estimates the deterministic $\tau_{r}$ in the presence of \gls{awgn} $n(t)$ and an unknown random constant amplitude of $s_{c, 2}(t)$.  Thus, the \gls{pdf} of $z(t)$ is conditioned on $\theta=\tau_{r}$ and $s_{c,2}(t)$. In the vector form, it is given by
$p\left(\mathbf{z};\theta,s_{c,2}\right) \sim \mathcal{CN}\left(a_{r}\sqrt{P_{r}}\,\mathbf{r}_{\tau_{r}}+b_{c}\sqrt{P_{c,2}}\,{s}_{c,2}\,\mathbf{h},\,\sigma_{n}^{2}\mathbf{I}\right),$ where $\mathbf{r}_{\tau_{r}}$ is the sampled radar pulse, ${s}_{c,2}$ is the unknown random constant amplitude and $\mathbf{h}$ is sampled unit-energy pulse. Next, we obtain  $p\left(\mathbf{z};\theta=\tau_{r}\right)$ by marginalizing $p\left(\mathbf{z};\theta,s_{c,2}\right)$  over $s_{c,2}$ with its distribution $p(s_{c,2})$ as \cite{NConv@IN}
\begin{equation}\label{eq:Radar_Sig_Dist_SN_P}
\begin{split}
    p\left(\mathbf{z};\theta=\tau_{r}\right)&= \int p\left(\mathbf{z};\theta,s_{c,2}\right)\,p\left(s_{c,2}\right)ds_{c,2},
\end{split}
\end{equation}
which is equivalent to a convolution of two Gaussians, giving:
\begin{equation}\label{eq:Radar_Sig_Dist_N}
    p\left(\mathbf{z};\theta\right) \sim \mathcal{CN}\left(a_{r}\sqrt{P_{r}}\,\mathbf{r}_{\tau_{r}}, \lvert{b_{c}}\rvert^{2}{P_{c,2}}\,\mathbf{h}\mathbf{h}^{H}+\sigma_{n}^{2}\mathbf{I}\right).
\end{equation}
Building on \eqref{eq:Radar_Sig_Dist_N}, with mean $\boldsymbol{u}_{z}(\tau_{r})=a_{r}\sqrt{P_{r}}\,\mathbf{r}_{\tau_{r}}$ and covariance matrix $\boldsymbol{\Sigma}_{z}=\lvert{b_{c}}\rvert^{2}{P_{c,2}}\,\mathbf{h}\mathbf{h}^{H}+\sigma_{n}^{2}\mathbf{I}$, we calculate the \gls{fim} as}
\AM{\begin{subequations}
\begin{align}
   J(\tau_{r}) &= \left( \frac{\partial \boldsymbol{\mu}_{z}(\tau_{r})}{\partial \tau_{r}} \right)^{H} \boldsymbol{\Sigma}_{z}^{-1} \left( \frac{\partial \boldsymbol{\mu}_{z}(\tau_{r})}{\partial \tau_{r}} \right),\label{eq:FIM_Expa}\\ 
   &=\lvert{a_{r}}\rvert^{2}P_{r}(\mathbf{r}_{\tau_{r}}^{'})^{H}(\lvert{b_{c}}\rvert^{2}{P_{c,2}}\,\mathbf{h}\mathbf{h}^{H}+\sigma_{n}^{2}\mathbf{I})^{-1}(\mathbf{r}_{\tau_{r}}^{'}),\label{eq:FIM_Expb}\\
   &=\frac{\lvert{a_{r}}\rvert^{2}P_{r}}{\sigma_{n}^{2}}\left(\lVert \smash{\mathbf{r}_{\tau_{r}}^{'}} \rVert^2 - \frac{\varrho \lvert{\mathbf{h}^{H}\mathbf{r}_{\tau_{r}}^{'}}\rvert^{2}}{1 + \varrho \|\mathbf{h}\|^2}\right),\label{eq:FIM_Expc}\\
&\geq\frac{{\lvert{a_{r}}\rvert^{2}P_{r}}\lVert \smash{\mathbf{r}_{\tau_{r}}^{'}} \rVert^2}{\sigma_{n}^{2} + \lvert{b_{c}}\rvert^{2}{P_{c,2}}}\label{eq:FIM_Expd}
\end{align}
\end{subequations}
where \eqref{eq:FIM_Expc} uses the Sherman–Morrison formula with $\varrho = \lvert b_c \rvert^2 P_{c,2} / \sigma_n^2$, and \eqref{eq:FIM_Expd} applies the Cauchy–Schwarz inequality. Notably, \eqref{eq:FIM_Expd} leads to a more pessimistic \gls{crlb} for the proposed \gls{rs}-inspired approach.}
 Subsequently, we  obtain the \gls{crlb} for the estimate of ${\tau_{r}}$ as \cite{Bliss@InnerBound}
\begin{equation}\label{eq:var_est}
\sigma_{\tau_{r},\textrm{est}}^{2}\geq\left(\frac{\sigma_{n}^{2} + \abs{b_{c}}^{2}P_{c,2}}{2\gamma^{2}B^{2}\left(TB\right)\abs{ a_{r}}^{2}P_{r}}\right).
\end{equation}
With the obtained \gls{crlb} for time-delay estimation, we calculate the ergodic \gls{reir}\cite{Yuanwei@Uplink_ISaC,Bliss@InnerBound,Yuanwei@NOMA_ISaC}. \AM{\gls{reir} is the \gls{dir} equivalent of the communication user, and quantifies the information gain as the difference between the entropy of the estimated parameter and its estimation uncertainty\cite{Bliss@InnerBound,Yuanwei@NOMA_ISaC,Yuanwei@Uplink_ISaC}, given by}
\begin{equation}\label{eq:R_est_Stan}
R_{\textrm{est}}\leq \frac{\delta}{2T}\log_{2}\left(1+\frac{\sigma_{\tau_{r},\textrm{proc}}^{2}}{\sigma_{\tau_{r},\textrm{est}}^{2}}\right),
\end{equation}
where $\delta$ is the radar duty factor, such that $T_{\textrm{pri}}=T/\delta$. Next, with the estimated time-delay, we suppress the radar signal from $z\left(t\right)$ and estimate the second communication stream. 
\vspace{-0.27cm}
\subsection{Second Communication Stream With Estimated Radar Return Suppressed}
The received signal at the communication receiver with estimated radar return suppressed can be written as
\begin{equation}
\begin{split}
    \Tilde{y}\left(t\right)=& b_{c}\sqrt{P_{c,2}}\,{s}_{c,2}\left(t\right)+\\
    &\sqrt{P_{r}}\,a_{r}\left[{r}\left(t-\tau_{r}\right)-{r}\left(t-\tau_{r,\textrm{est}}\right)\right]+n\left(t\right),     
\end{split}
\end{equation}
where we now have $\tau_{r}=\tau_{r,\textrm{est}}+n_{\tau_{r},\textrm{est}}$, similar to the predicted radar return. Moreover, the lower bound on the variance of $n_{\tau_{r},\textrm{est}}$ is obtained from equation \eqref{eq:var_est}. Subsequently, the interference plus noise power for ${s}_{c,2}$ is given by
\begin{equation}\label{eq:noise_power_sc2}
     \sigma_{\textrm{int}+\textrm{n},2}^{2} = P_{r}\abs{a_{r}}^2\gamma^2 B^2 \sigma_{\tau_{r},\textrm{est}}^{2} +\sigma_{n}^{2}. 
\end{equation}
Using \eqref{eq:noise_power_sc2}, we can calculate the \gls{dir} for the second communication stream. Following this, we calculate the performance bounds with \gls{rs} for both sensing and communication.
\vspace{-0.1cm}
\section{Performance bounds}
As mentioned earlier, we consider ergodic \gls{reir} for the radar target and ergodic \gls{dir} for the communication user as performance measures. To this end, the bounds are derived assuming that the radar pulse duration $T$ is held constant\cite{Bliss@InnerBound}.  Subsequently, using equations \eqref{eq:var_est} and \eqref{eq:R_est_Stan}, the ergodic \gls{reir} bound for the target is given by 
\begin{equation}\label{eq:R_est_InnerBound}
    R_{\textrm{est}}^{\textrm{RS}} \leq \frac{\delta}{2T}\log_{2}\left(1+\frac{2\sigma_{\tau_{r},\textrm{proc}}^{2}\gamma^{2}B^{2}\left(TB\right)\abs{a_{r}}^{2}P_{r}}{\sigma_{n}^2+\abs{b_{c}}^{2}P_{c,2}}\right).
\end{equation}
Next, the bound on the \gls{dir} of the communication user is given by  $R_{c}^{\textrm{RS}}\leq R_{c,1}+R_{c,2}$, where $R_{c,1}$ and  $R_{c,2}$ are the data rates of communication streams $1$ and $2$, respectively, expressed as
\begin{align}\label{eq:RS_Comm_Rates}
    R_{c,1}\leq& B\log_{2}\left(1+\frac{\abs{b_{c}}^{2}P_{c,1}}{\sigma_{n}^2+P_{r}\abs{a_{r}}^{2}\gamma^{2}B^{2}\sigma_{\tau_{r},\textrm{proc}}^{2}+\abs{b_{c}}^{2}P_{c,2}}\right)\nonumber\\
    R_{c,2}\leq& B\log_{2}\left(1+\frac{\abs{b_{c}}^{2}P_{c,2}}{\sigma_{n}^2+P_{r}\abs{a_{r}}^{2}\gamma^{2}B^{2}\sigma_{\tau_{r},\textrm{est}}^{2}}\right).
\end{align}
The inner bounds of the communication user can be obtained by varying the power split between $P_{c,1}$ and $P_{c,2}$ as $P_{c,1}=(1-\alpha)P_{c}$ and $P_{c,2}=\alpha P_{c}$, where $\alpha \in [0,1]$. Next, we calculate the optimal $\alpha$ for which the communication user achieves the maximum ergodic \gls{dir}. \AM{The optimal power split, $\alpha_{\textrm{max}}^{\textrm{RS}}$, can be derived as $ \argmax_{\alpha \in [0,1]}\,R_{c,1}+R_{c,2}$. To obtain $\alpha_{\textrm{max}}^{\textrm{RS}}$, we set the derivative of $R_{c,1}+R_{c,2}$ with respect to $\alpha$ equal to zero. Subsequently, we get
\begin{align}\label{eq:quadratic_alpha}
{P_{bc}^{3}}\alpha^{2} + &{2P_{bc}^{2}\sigma_{n}^2}\alpha + {\sigma_{n}^{4}P_{bc}}\nonumber \\&-2\sigma_{n}^{2}P_{bc}P_{r}\abs{a_{r}}^{2}\gamma^{2}B^{2}\sigma_{\tau_{r},\textrm{proc}}^{2}TB=0,
\end{align}
where $P_{bc}=\abs{b_{c}}^{2}P_{c}$. Solving \eqref{eq:quadratic_alpha}, we get optimal $\alpha$ as
\begin{equation}\label{eq:opt_alpha}
    \alpha_{\textrm{max}}^{\textrm{RS}}= \frac{-\sigma_{n}^2+\abs{a_{r}}\gamma B \sigma_{\tau_r,\textrm{proc}}\sqrt{2P_{r}TB\sigma_{n}^2}}{\abs{b_{c}}^{2}P_{c}}.
\end{equation}
Although $\alpha_{\textrm{max}}^{\textrm{RS}}$ is real-valued, whether it lies within the interval $(0,1)$ depends on the system parameters in \eqref{eq:opt_alpha}. If it falls outside this range, it implies that one functionality’s signal should be decoded while fully treating the other as noise, effectively setting either $P_{c,1}$ or $P_{c,2}$ to zero.}
\vspace{-0.4cm}
\subsection{Inner Bounds on Performance with Baseline Schemes}
For comparison, we also delineate the inner bounds for \gls{oma}-inspired (spectral isolation) and \gls{noma}-inspired (resource sharing with \gls{sic} only) approaches derived from \cite{Bliss@InnerBound,Yuanwei@NOMA_ISaC,Yuanwei@Uplink_ISaC}. Following this, the performance inner bounds for \gls{oma}-inspired approach are given by\cite{Bliss@InnerBound}
\begin{align}
    R_{\textrm{est}}^{\textrm{\gls{oma}}} \leq &\,\frac{\delta}{2T}\log_{2}\left(1+\frac{2\sigma_{\tau_{r},\textrm{proc}}^{2}\gamma^{2}\left(1-\mu\right)^{2}B^{2}TB\abs{a_{r}}^{2}P_{r}}{\sigma_{n}^{2}}\right), \nonumber\\
    R_{\textrm{c}}^{\textrm{OMA}} \leq &\,\mu B\log_{2}\left(1+\frac{\abs{b_{c}}^{2}P_{c}}{\mu \sigma_{n}^{2}}\right),
\end{align}
where $\mu\in[0,1]$ partitions the total bandwidth $B$ into two sub-bands, one for the radar target, $B_{r}=(1-\mu)B$, and the other for the communication user, $B_{c}=\mu B$ \cite{Bliss@InnerBound}. With this approach, each functionality operates without interference from the other. 
\par Next, for the \gls{noma}-inspired approach, the inner bounds are obtained by considering decoding of the communication user's signal first and removing it using \gls{sic}, \AM{following\cite{Yuanwei@NOMA_ISaC,Bliss@InnerBound}.} Subsequently, the time-delay estimation is done without any interference from the communication user. Consequently,  the inner bounds for \gls{noma}-inspired approach are given by
\begin{align}\label{eq:sic_innerbound}
    R_{\textrm{est}}^{\textrm{\gls{noma}}} \leq & \frac{\delta}{2T}\log_{2}\left(1+\frac{2\sigma_{\tau_{r},\textrm{proc}}^{2}\gamma^{2}B^2\left(TB\right)\abs{a_{r}}^{2}P_{r}}{\sigma_{n}^{2}}\right),\nonumber\\
    R_{\textrm{c}}^{\textrm{\gls{noma}}}\leq & B\log_{2}\left(1+\frac{\abs{b_{c}}^{2}P_{c}}{\sigma_{n}^{2}+P_{r}\abs{a_{r}}^{2}\gamma^{2}B^{2}\sigma_{\tau_{r},\textrm{proc}}^{2}}\right).
\end{align}
\AM{It is worth noting that \eqref{eq:sic_innerbound} can be obtained from \eqref{eq:R_est_InnerBound} and \eqref{eq:RS_Comm_Rates} as well, by setting $P_{c,1} = P_{c}$ and $P_{c,2} = 0$.}
\vspace{-0.4cm}
\section{Results}\label{NumRes}
 \AM{In this section, we demonstrate the derived inner bounds for the proposed \gls{rs}-inspired approach}. To this end, the parameters considered are delineated in Table \ref{tab:Table_params}\cite{Bliss@InnerBound}. We assume that the signal of the communication user is received through an antenna sidelobe, leading to different radar and communications receive signal gains. Together with the antenna gains, we calculate large scale coefficients in ${a_{r}}$ using \cite[eq. $(2.8)$]{richards2010principles}, and ${b_{c}}$  using the standard free-space path-loss model. \AM{Small scale fading is modeled as Rayleigh distribution}. Moreover, $\sigma_{\tau_{r},\textrm{proc}}$ is calculated by dividing the target process standard deviation by the speed of light. {We adopt the parameters and path-loss models from \cite{Bliss@InnerBound} to ensure fair comparison with baseline approaches. The results generalize to other cellular scenarios by proportionally scaling the parameters in Table \ref{tab:Table_params}}. Finally, the bounds illustrated in this section are obtained by producing the convex hull of all contributing inner bounds \cite{Bliss@InnerBound}.
 \begin{table}[!t]
	\caption{Simulation parameters\cite{Bliss@InnerBound}}\vspace{-0.1cm}
	    \label{tab:Table_params}\centering
	\begin{tabular}{l l l}
		\toprule[0.4mm]
		\textbf{Parameter} & \textbf{Value}\\
  		\toprule[0.4mm]
   Bandwidth $\left(B\right)$  & $5$ MHz\\
   Frequency $\left(f\right)$ &  $3$ GHz\\
   Effective Temperature &  $1000$K \\
   Communications range &  $10$ km\\
   Communications power $\left(P_{c}\right)$ & $100$ W\\
   Communications antenna gain & $0$ dBi\\
   Communications receiver side-lobe gain &  $10$ dBi\\
   Radar target range &  $100$ Km\\
   Radar antenna gain &  $30$ dBi\\
   Radar Power $\left(P_{r}\right)$  &  $100$ kW\\
   Target cross section & $10\,\textrm{m}^{2}$ \\
   Target process standard deviation & $100$\,m\\
   Time-Bandwidth product $\left(TB\right)$ &  $100$\\
   Radar duty factor $\left(\delta\right)$ & $0.01$\\
    \bottomrule[0.4mm]	
 \end{tabular}\vspace{-0.5cm}
\end{table}
\par
\begin{figure}[!t]
    \centering
    \includegraphics[width=0.7\linewidth]{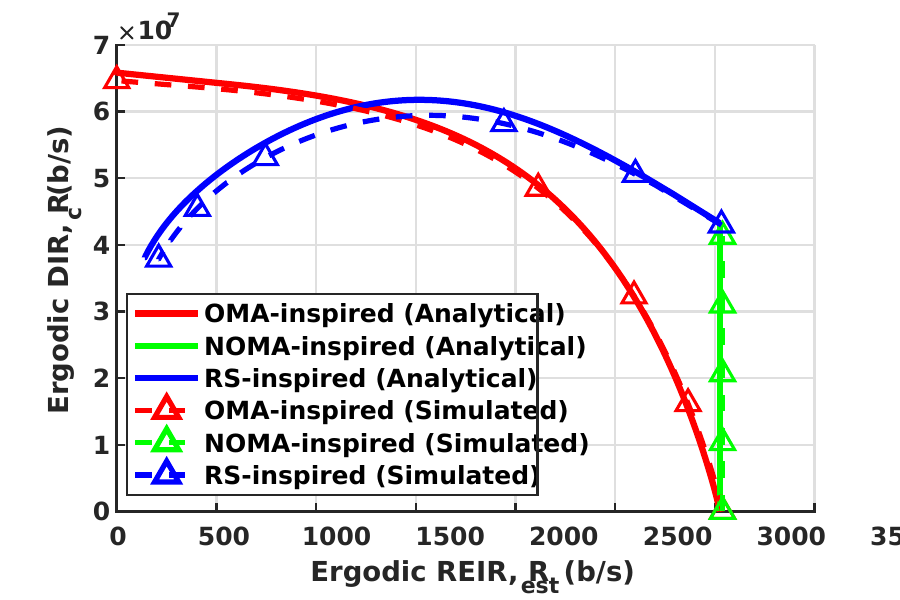}
    \caption{Inner bounds of multiple access-inspired schemes. }
    \label{fig:Inner_Bound}\vspace{-0.6cm}
\end{figure}

 {Fig.~\ref{fig:Inner_Bound}
illustrates the derived inner bounds and simulations results on the performance of the joint sensing-communication system, with the proposed scheme labeled as \gls{rs}-inspired.} Baseline schemes are labeled as \gls{oma}-inspired and \gls{noma}-inspired \cite{Bliss@InnerBound,Yuanwei@NOMA_ISaC}. We begin with the baseline schemes. In the \gls{oma}-inspired approach, as $\mu$ increases from 0 to 1, the ergodic \gls{reir} decreases while the communication user’s \gls{dir} increases, due to greater bandwidth allocation. The \gls{noma}-inspired inner bounds are shown as a green vertical line, where each point corresponds to increasing the communication user’s transmit power from $0$ to $P_{c}$. The maximum ergodic \gls{dir}, achieved at full power $P_{c}$, corresponds to the expression in~\eqref{eq:sic_innerbound}. Since time-delay estimation is performed after \gls{sic} of the communication signal, the \gls{reir} remains unaffected by the communication user's power. However, the communication user remains interference-limited due to the radar signal. {The simulation results for both the \gls{oma}- and \gls{noma}-inspired approaches closely align with their respective bounds.}
\par \AM{Turning to the \gls{rs} scheme, the inner bounds are obtained by varying $\alpha$ from 0 to 1. At $\alpha = 0$, we have $P_{c,2} = 0$, and the \gls{rs}-inspired curve begins at the upper vertex of the \gls{noma}-inspired vertical line, as both schemes yield identical ergodic performance; see~\eqref{eq:R_est_InnerBound}, \eqref{eq:RS_Comm_Rates}, and \eqref{eq:sic_innerbound}}. As $\alpha$ increases, the radar’s ergodic \gls{reir} decreases, while the communication user’s ergodic \gls{dir} improves. This is due to the increase in $P_{c,2}$, which raises the effective noise $\sigma_{\textrm{eff}}^{2}$ for radar sensing, thereby increasing the \gls{crlb} and degrading the estimation accuracy of the radar parameter $\tau_{r}$. In contrast, the communication user benefits from the flexibility of the \gls{rs} scheme. Message splitting allows the two streams to experience different interference conditions: $s_{c,1}$ is affected by both the predicted radar return suppressed signal and $s_{c,2}$, while $s_{c,2}$ is only interfered by the estimated radar return suppressed signal. As $P_{c,2}$ increases (i.e., $\alpha$ grows), the total ergodic \gls{dir} initially improves. However, beyond a certain point, the rate of $s_{c,1}$ drops significantly, \gls{reir} continues to degrade, and increased interference also begins to impair the rate of $s_{c,2}$. Consequently, the total ergodic \gls{dir} starts to decline beyond a critical value of $\alpha$. {The simulation results for the \gls{rs}-inspired approach closely follow the derived bounds, though with a slightly larger gap than the baselines. This can be attributed to the summation of two rate expressions for \gls{rs} in \eqref{eq:R_est_InnerBound}, as opposed to one in baseline schemes.}
\begin{figure}[!b]
    \vspace{-0.8cm}\centering
    \includegraphics[width=0.6\linewidth]{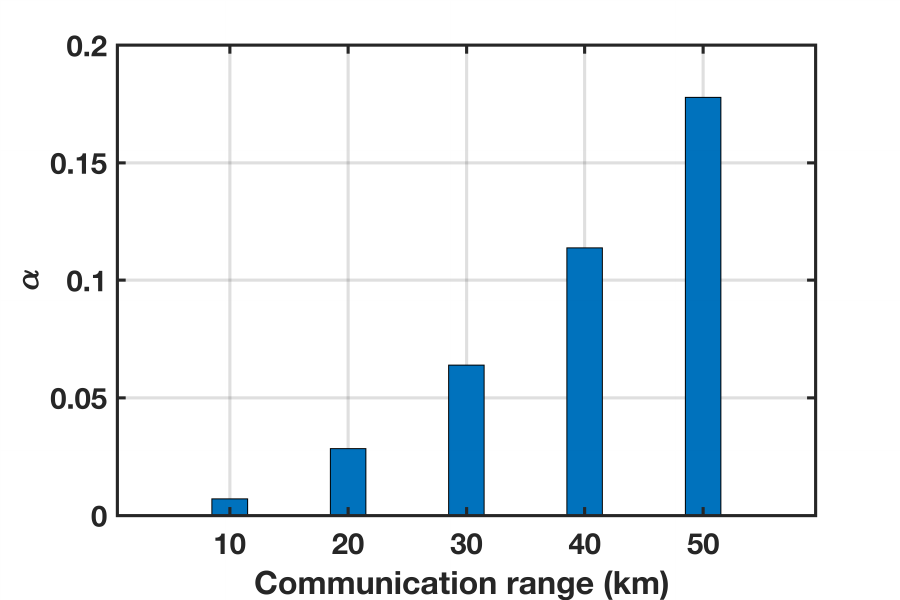}
    \caption{{Value of $\alpha$ with increasing communication range.}}
    \label{fig:Alpha_plot}
\end{figure}
\par We determine the point of inflection of the \gls{rs} curve using equation~\eqref{eq:opt_alpha}. For the parameters listed in Table~\ref{tab:Table_params}, the optimal value is $\alpha_{\textrm{max}}^{\textrm{RS}} = 0.0071$. This value is verified by locating the $\alpha$ corresponding to the maximum point on the \gls{rs} curve in Fig.~\ref{fig:Inner_Bound}. The small value of $\alpha_{\textrm{max}}^{\textrm{RS}}$ arises from the communication user being significantly closer to the \gls{bs} than the radar target. As a result, the communication user's received signal power is substantially higher than the radar echo, leading to minimal interference from the radar on the first communication stream. Consequently, only a small portion of power needs to be allocated to the second stream. {This can be observed in Fig.~\ref{fig:Alpha_plot}.}
\par \AM{Finally, the left endpoint of the \gls{rs} curve corresponds to the case $P_{c,2} = P_{c}$ where sensing is performed first while treating the communication signal as interference. This results in degraded performance for both functionalities under the given parameters. Note that this point also corresponds to the \gls{noma}-inspired approach, if the sensing signal were always decoded first.} Nonetheless, the \gls{rs} scheme achieves larger inner bounds than the \gls{oma}- and \gls{noma}-inspired approaches, enabling a better trade-off up to a certain power split.
\vspace{-0.1cm}
\section{Conclusion}\label{Concl}
In this paper, we proposed a novel approach utilizing \gls{rs} at the communication user and \gls{sic} at the \gls{bs} to establish joint sensing-communication performance inner bounds. \AM{Using ergodic \gls{reir} for radar and ergodic \gls{dir} for communication as performance metrics, we analyzed the \gls{rs}-inspired approach against baseline schemes; \gls{oma} and \gls{noma}-inspired approaches. Our results demonstrate that \gls{rs} effectively mitigates inter-functionality interference and achieves a larger inner bound up to a certain power split, thereby allowing for an improved spectral efficiency of the system}. Additionally, we derived a closed-form expression for the optimal power split that maximizes the ergodic \gls{dir}. While traditional \gls{rs} applies to digital signals, our work is the first work to extend it as a general framework for non-orthogonal sensing and communication waveforms. 

\ifCLASSOPTIONcaptionsoff
  \newpage
\fi

\bibliographystyle{IEEEtran}
\bibliography{reference}
\end{document}